%% file: main.tex
\documentclass[aps,prl,reprint,superscriptaddress,longbibliography,raggedbottom,nofootinbib,nobibnotes]{revtex4-2}
\usepackage[T1]{fontenc}
\usepackage{amsmath,amssymb,booktabs,graphicx,microtype,needspace}
\usepackage[colorlinks=true,linkcolor=blue,citecolor=blue,urlcolor=blue]{hyperref}
\hypersetup{pdftitle={Next-to-next-to-leading-order QCD corrections to radiative and hadronic Upsilon(1S) decays}}
\newcommand{\MSbar}{\overline{\mathrm{MS}}}
\newcommand{\dd}{\mathrm d}
\input{window_paper_data.tex}
\allowdisplaybreaks[1]
\begin{document}
\title{Next-to-next-to-leading-order QCD corrections to radiative and hadronic \texorpdfstring{$\Upsilon(1S)$}{Upsilon(1S)} decays}
\author{Xiang-Yu Chang}
\thanks{These authors contributed equally to this work and share first authorship.}
\author{Chao Zeng}
\thanks{These authors contributed equally to this work and share first authorship.}
\author{Ruichang Niu}
\author{Bin Gong}
\email[Contact author: ]{twain@ihep.ac.cn}
\author{Jian-Xiong Wang}
\email[Contact author: ]{jxwang@ihep.ac.cn}
\affiliation{Institute of High Energy Physics, Chinese Academy of Sciences, Beijing 100049, China}
\affiliation{University of Chinese Academy of Sciences, Beijing 100049, China}
\hypersetup{pdfauthor={Xiang-Yu Chang, Chao Zeng, Ruichang Niu, Bin Gong, Jian-Xiong Wang}}
\begin{abstract}
We compute complete next-to-next-to-leading-order (NNLO) QCD corrections to
radiative and inclusive hadronic $\Upsilon(1S)$ decays in nonrelativistic QCD
at leading order in velocity.
At $\mu_R=M_\Upsilon$, they lower the direct next-to-leading-order (NLO)
radiative width in $0.65\leq 2E_\gamma/M_\Upsilon\leq0.92$ by $41\%$ and the
NLO hadronic width by $32\%$, improving agreement with the CLEO spectrum and
measured hadronic width.
The width ratio cancels the nonperturbative matrix element and suppresses scale
dependence in its NNLO expansion, giving
$\alpha_s^{(5)}(M_Z)=\FitExpandedHighWithError$ from CLEO data, consistent with
the world average.
\end{abstract}
\maketitle

\Needspace*{3\baselineskip}
\textit{Introduction.---}
Heavy-quarkonium annihilation decays cleanly probe QCD at intermediate scales
\cite{QuarkoniumReview,PDG}.
In nonrelativistic QCD (NRQCD), decay rates factorize into perturbative
short-distance coefficients in $\alpha_s$ and nonperturbative long-distance
matrix elements (LDMEs) ordered in the heavy-quark velocity $v$
\cite{BBL,QuarkoniumEFT}.
Bottomonium is well suited for precision tests, with $m_b\approx4.7$~GeV in
the perturbative regime and small relativistic effects ($v_b^2\sim0.1$).

In $\Upsilon(1S)$ annihilation, inclusive hadronic decay
($\Upsilon\to\allowbreak ggg$) dominates the width, while radiative decay
($\Upsilon\to\allowbreak\gamma gg$) probes the photon spectrum and, via their
width ratio, determines $\alpha_s$ free of the leading LDME
\cite{CLEO2006,Garcia2005,RadiativeAlpha2007}.
Next-to-leading-order (NLO) QCD corrections to both channels have long been
known \cite{MackenzieLepage,Kraemer1999}, yet remain insufficient: for standard
inputs at $\mu_R=\allowbreak M_\Upsilon$, the NLO hadronic width exceeds the
measured total $\Upsilon(1S)$ width \cite{MackenzieLepage,PDG}, while the NLO
radiative calculation includes only direct $b$-quark emission \cite{Kraemer1999}
and missing higher-order QCD corrections limit spectrum descriptions and
$\alpha_s$ determinations
\cite{CLEO2006,Garcia2005,RadiativeAlpha2007,SotoProspects}.
Meanwhile, next-to-next-to-leading-order (NNLO) QCD calculations have been
extended to many two-body quarkonium processes
\cite{EtaNNLO,SangChicGamma2020,YuGammaEtac2021,SangJpsiChic2023,HuangGongWang,SangZQuarkonium2023,SangDoubleJpsi2023,ZhangBottomoniumDoubleJpsi2023,HuangDoubleJpsi2024,FengJpsiEtac2024,ChenJpsiEtac2026,SoftImprovedMatching}
and recently to three-body channels with complete $J/\psi\to\allowbreak 3\gamma$
\cite{ThreePhoton} and approximate $e^+e^-\to\allowbreak J/\psi+\allowbreak c\bar c$
\cite{ChangJpsiOpenCharm2026} results.
Two-loop NLO QCD corrections have also been computed for
$\eta_b\to\allowbreak J/\psi+\allowbreak\gamma$ \cite{YangEtabJpsiGamma2026}.
Large NNLO corrections in Refs.~\cite{EtaNNLO,ThreePhoton} further motivate
computing both $\Upsilon(1S)$ channels at NNLO.

Here we compute complete NNLO QCD corrections to radiative and inclusive
hadronic $\Upsilon(1S)$ decays in NRQCD at leading order in $v$, obtaining the
photon spectrum (including light-quark emission and fragmentation), partial
radiative width, and hadronic width.
Large negative NNLO corrections improve agreement with CLEO data and resolve
the NLO excess in the hadronic width, while the expanded window ratio cancels
the leading LDME and suppresses scale dependence to determine $\alpha_s$.

\Needspace*{3\baselineskip}
\textit{Calculation.---}
At leading order in $v$, kinematics fixes $m_b=M_\Upsilon/2$ with
$M_\Upsilon=\UpsilonMass$~GeV, and differential radiative and hadronic widths
factorize as
\begin{equation}
 \dd\Gamma_{\gamma/h}=\frac{4\,\dd F_{\gamma/h}}{M_\Upsilon^2}
 \langle O_1({}^3S_1;\mu_\Lambda)\rangle_\Upsilon
 +\mathcal O(v^2),
 \label{eq:factorization}
\end{equation}
with differential short-distance coefficient $\dd F_{\gamma/h}$ and
color-singlet LDME $\langle O_1\rangle_\Upsilon=3|R_b(0)|^2/(2\pi)$ at scale
$\mu_\Lambda$.

We work in conventional dimensional regularization with four massless flavors
($u,d,s,c$) and an on-shell $b$ quark, renormalizing $\alpha_s^{(5)}$ and
$\langle O_1\rangle_\Upsilon$ in the modified minimal subtraction ($\MSbar$)
scheme.
Amplitudes are generated and simplified with \textsc{FeynArts} and
\textsc{FORM} \cite{FeynArts,FORM}; loop integrals are reduced via Kira~3.1
\cite{KiraOriginal,Kira,Kira3} and SecRed \cite{SecRed} and evaluated with
AMFlow \cite{AMFlow}.
Real-emission infrared singularities are subtracted using local analytic sector
\cite{LASS,LASSIntegrals,LASSComplete} and nested soft-collinear
\cite{NSC,NSCDecays} schemes.
After UV renormalization and NRQCD matching
\cite{Catani,BDK,LeptonicBSS,ChungWavefunctions}, QCD infrared poles cancel
analytically, while residual $q\parallel\gamma$ QED collinear poles are
absorbed into photon fragmentation functions below.
Two-loop--tree interference is evaluated by Gauss--Kronrod quadrature and
subtracted real emission by Monte Carlo (MC) integration; details will be
presented elsewhere.

For the partial radiative width in a photon-energy window (at fixed
fragmentation input) or inclusive hadronic width, $F_{\gamma/h}$ expands around
its leading-order (LO) value $F_{\gamma/h}^{(0)}$ in
$\alpha_s=\alpha_s^{(5)}(\mu_R)$ as
\begin{equation}
 F_{\gamma/h}=F_{\gamma/h}^{(0)}\biggl[1
 +c_{\gamma/h}^{(1)}\frac{\alpha_s}{\pi}
 +c_{\gamma/h}^{(2)}\left(\frac{\alpha_s}{\pi}\right)^2
 +\mathcal O(\alpha_s^3)\biggr].
 \label{eq:short-distance}
\end{equation}%
\begin{figure*}[t]
\centering
\includegraphics[width=.98\textwidth]{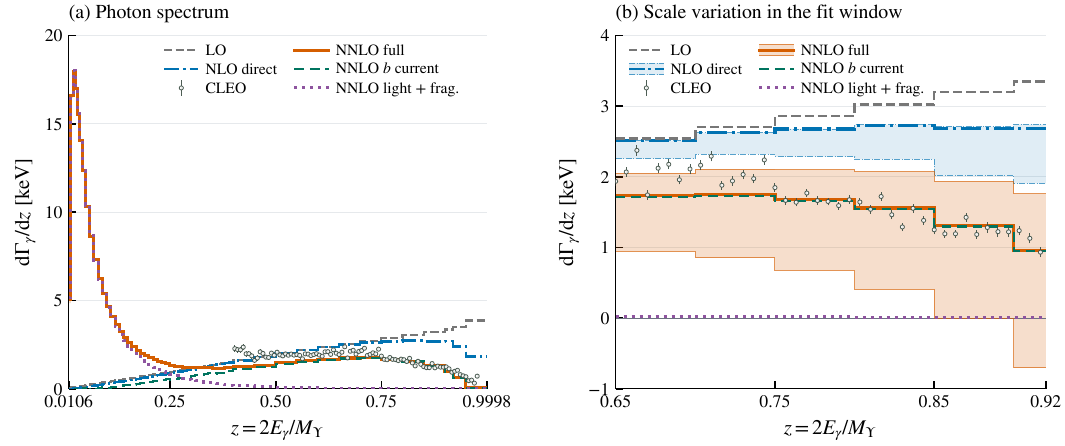}
\caption{Radiative $\Upsilon(1S)$ photon spectrum compared with reconstructed
CLEO data \cite{CLEO2006}.
(a) Central spectra ($\mu_R=\mu_F=M_\Upsilon$) from $E_\gamma=50$~MeV to the
single-pion threshold.
(b) Fit window $W=[0.65,0.92]$, with shaded $\mu_R\in[M_\Upsilon/2,2M_\Upsilon]$
bands for direct NLO and full NNLO.
Both panels also show the central NNLO $b$-current and light-quark plus
fragmentation contributions.}
\label{fig:spectrum}
\end{figure*}%
Renormalization group invariance and NRQCD matching determine the scale
dependence of the first- and second-order coefficients
\cite{LeptonicBSS,ChungWavefunctions}:
\begin{align}
 c_{\gamma/h}^{(1)}={}&\bar c_{\gamma/h}^{(1)}+\frac{p\beta_0}{4}L_R,\notag\\*
 c_{\gamma/h}^{(2)}={}&\bar c_{\gamma/h}^{(2)}
 +\left[\frac{(p+1)\beta_0}{4}\bar c_{\gamma/h}^{(1)}
 +\frac{p\beta_1}{16}\right]L_R\notag\\*
 &+\frac{p(p+1)\beta_0^2}{32}L_R^2
 -\frac{70\pi^2}{27}L_\Lambda ,
 \label{eq:scale-coefficients}
\end{align}
where barred quantities correspond to $\mu_R=M_\Upsilon$ and
$\mu_\Lambda=M_\Upsilon/2$, $L_R=\ln(\mu_R^2/M_\Upsilon^2)$,
$L_\Lambda=\ln(4\mu_\Lambda^2/M_\Upsilon^2)$, $\beta_0=23/3$, $\beta_1=116/3$,
and the power $p$ is $2$ for radiative decay and $3$ for hadronic decay.
For central predictions we use three-loop running with the Particle Data Group
(PDG) world average $\alpha_s^{(5)}(M_Z)=\allowbreak 0.1180$ \cite{PDG},
$|R_b(0)|^2=\allowbreak 6.477$~GeV$^3$ \cite{EichtenQuigg,ThreePhoton}, and the
Thomson coupling $\alpha(0)^{-1}\simeq\allowbreak\InverseAlphaZero$
\cite{CODATA2022}, varying $\mu_R\in\allowbreak[M_\Upsilon/2,2M_\Upsilon]$ at
fixed $\mu_\Lambda=\allowbreak M_\Upsilon/2$.

\textit{Radiative decay.---}
In the $\Upsilon(1S)$ rest frame ($z=\allowbreak 2E_\gamma/M_\Upsilon$),
partonic amplitudes through NLO ($\mathcal O(\alpha\alpha_s^3)$) couple the
photon only to the $b$ quark, giving infrared-finite LO and direct NLO spectra
without fragmentation \cite{Kraemer1999}, which we use as our LO and NLO
benchmarks.
At NNLO ($\mathcal O(\alpha\alpha_s^4)$), massless light quarks also radiate
photons: bottom--light interference---between amplitudes with the photon
coupled to the $b$ quark and to a light quark---is infrared finite, whereas
squared light-quark emission in $\Upsilon\to\allowbreak\gamma+gg+q\bar q$
produces collinear $q\parallel\gamma$ singularities.
Factorizing these singularities in the $\MSbar$ scheme at scale $\mu_F$ into
photon fragmentation functions $D_{i\to\gamma}$
\cite{CataniHautmann,GehrmannSchurmann2022} gives, for $z>0$,
\begin{equation}
 \frac{\dd\Gamma_\gamma}{\dd z}
 =C_\gamma(z;\mu_F)
 +\sum_i\int_z^1\frac{\dd y}{y}\,
 C_i(y;\mu_F)D_{i\to\gamma}(z/y;\mu_F),
 \label{eq:photon-spectrum}
\end{equation}
where $C_\gamma$ is the subtracted direct spectrum and $C_i$ ($i=g,q,\bar q$)
are inclusive parton spectra.
With $D_{i\to\gamma}=\allowbreak\mathcal O(\alpha)$
\cite{GehrmannSchurmann2022}, our full NNLO prediction evaluates
Eq.~\eqref{eq:photon-spectrum} through $\mathcal O(\alpha\alpha_s^4)$,
combining $C_\gamma$ through $\mathcal O(\alpha\alpha_s^4)$ with the
$\mathcal O(\alpha_s^3)$ and $\mathcal O(\alpha_s^4)$ parton spectra $C_i$ (so
fragmentation enters via $C_g\otimes\allowbreak D_{g\to\gamma}$ at
$\mathcal O(\alpha\alpha_s^3)$, omitted from direct NLO, and
$C_i\otimes\allowbreak D_{i\to\gamma}$ at $\mathcal O(\alpha\alpha_s^4)$).
We evolve $D_{i\to\gamma}$ to $\mu_F=M_\Upsilon$ at four-flavor NLO using
Bourhis--Fontannaz--Guillet (BFG) set~II \cite{BFGPhoton}.

Figure~\ref{fig:spectrum}(a) displays the central spectrum from
$E_\gamma=\allowbreak 50$~MeV ($z\simeq\allowbreak 0.0106$) to the single-pion
threshold $m_X=\allowbreak M_\Upsilon\sqrt{1-z}=\allowbreak m_{\pi^0}$
($z\simeq\allowbreak 0.9998$), comparing LO and direct NLO with full NNLO and
its $b$-current and light-quark plus fragmentation contributions.
For emission from the $b$ current, higher-order QCD corrections successively
reduce the spectrum from LO to direct NLO and NNLO, and all three curves remain
small at low $z$.
By contrast, the NNLO light-quark plus fragmentation contribution is small at
moderate and large $z$, but grows rapidly below $z\lesssim0.3$ due to soft
$\dd z/z$ bremsstrahlung and fragmentation, reaching $\sim18$~keV near
$z\approx0.025$ before negative $\mathcal O(\alpha\alpha_s^4)$ fragmentation
corrections reduce the lowest bin to $\sim5$~keV.
Together with large small-$z$ timelike logarithms \cite{BFGPhoton}, this
behavior invalidates fixed-order predictions as $z\to 0$.

To compare with experiment, we reconstruct the CLEO spectrum over
$0.40\leq\allowbreak z\leq\allowbreak0.98$ from Fig.~15 of Ref.~\cite{CLEO2006}
using the quoted efficiency, an acceptance $A\simeq 0.67$ \cite{CLEO1997}, and
the PDG total width \cite{PDG}.
At central scales, the NNLO prediction lies below the data for
$0.40\leq\allowbreak z\leq\allowbreak0.75$ and $z\gtrsim0.90$, and agrees well
with them for $0.75\lesssim\allowbreak z\leq\allowbreak0.90$
[Fig.~\ref{fig:spectrum}(a)].

To avoid the low-$z$ fragmentation region and suppress $z\to 1$ endpoint
effects, quantitative comparisons are restricted to the window
$W=[0.65,0.92]$, where CLEO fitted the Garcia i Tormo--Soto (GS) model
\cite{CLEO2006,Garcia2005}.
\begin{figure*}[t]
\centering
\includegraphics[width=.98\textwidth]{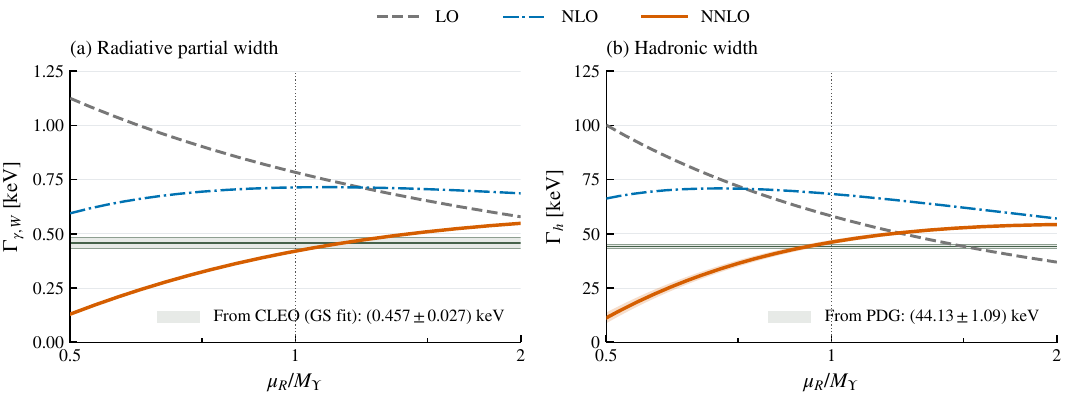}
\caption{Scale dependence of (a) $\Gamma_{\gamma,W}$ (direct at NLO) and (b)
$\Gamma_h$ for $\alpha_s(M_\Upsilon)=\ReferenceAlpha$.
Narrow bands on the NNLO curves show numerical uncertainties, and horizontal
bands show the experimental estimates.}
\label{fig:widths}
\end{figure*}
In this window [Fig.~\ref{fig:spectrum}(b)], higher-order corrections from LO
to direct NLO and full NNLO bring the theoretical spectrum closer to the CLEO
data, which lie mostly within the NNLO scale band, while central NNLO nearly
coincides with its $b$-current contribution.
For small $\mu_R$, the NNLO spectrum becomes negative in the highest-$z$ bins
of $W$, reflecting large $z\to 1$ endpoint logarithms that require resummation
\cite{Garcia2005}.

Integrating Eq.~\eqref{eq:photon-spectrum} over $W$ defines the partial
radiative width
\begin{equation}
 \Gamma_{\gamma,W}=\int_W\dd z\,
       \frac{\dd\Gamma_\gamma}{\dd z},
 \qquad W=[0.65,0.92],
 \label{eq:window}
\end{equation}
with central coefficients
\begin{equation}
 \begin{aligned}
 \bar c_{\gamma,W,\mathrm{dir}}^{(1)}&=\WindowCOne\pm\WindowCOneError_{\rm MC},\\
 \bar c_{\gamma,W}^{(2)}&=\WindowCTwo\pm\WindowCTwoError_{\rm MC},
 \end{aligned}
 \label{eq:radiative-coefficient}
\end{equation}
and reference widths
\begin{equation}
 \begin{aligned}
 \Gamma_{\gamma,W}^{\rm LO}&=\WindowWidthLO\ {\rm keV},\\
 \Gamma_{\gamma,W}^{\rm NLO,dir}&=(\WindowWidthNLO
                  \pm\WindowWidthNLOError_{\rm MC})\ {\rm keV},\\
 \Gamma_{\gamma,W}^{\rm NNLO}&=(\WindowWidthNNLO
                  \pm\WindowWidthError_{\rm MC})\ {\rm keV},
 \end{aligned}
 \label{eq:window-widths}
\end{equation}
where $\Gamma_{\gamma,W}^{\rm NNLO}$ also includes the $\mathcal
O(\alpha\alpha_s^3)$ $C_g\otimes\allowbreak D_{g\to\gamma}$ contribution from
Eq.~\eqref{eq:photon-spectrum}.
Contributions beyond the squared $b$ current---bottom--light interference
($\ExtraInterferencePercent\%$) and squared light-quark emission plus
fragmentation ($\ExtraSquarePercent\%$)---account for only
$\ExtraTotalPercent\%$ of $\Gamma_{\gamma,W}^{\rm NNLO}$ at central scales
[Fig.~\ref{fig:spectrum}(b)].
At $\mu_R=M_\Upsilon$, $\Gamma_{\gamma,W}^{\rm NNLO}$ is
$\WindowNNLOReduction\%$ below $\Gamma_{\gamma,W}^{\rm NLO,dir}$ and close to
the experimental estimate $(0.457\pm0.027)$~keV inferred from CLEO's GS fit and
$\Gamma_h^{\rm PDG}$ below [Fig.~\ref{fig:widths}(a)].
For $\mu_R\in\allowbreak[M_\Upsilon/2,2M_\Upsilon]$ [Fig.~\ref{fig:widths}(a)],
$\Gamma_{\gamma,W}^{\rm NNLO}$ increases from $\WindowNNLOMinimum$ to
$\WindowNNLOMaximum$~keV (opposite to LO), corresponding to
$-\WindowNNLOScaleDown\%/+\WindowNNLOScaleUp\%$ versus
$-\WindowNLOScaleDown\%/+\WindowNLOScaleUp\%$ at direct NLO, which has a
maximum near $M_\Upsilon$.

\begin{figure*}[t]
\centering
\includegraphics[width=.98\textwidth]{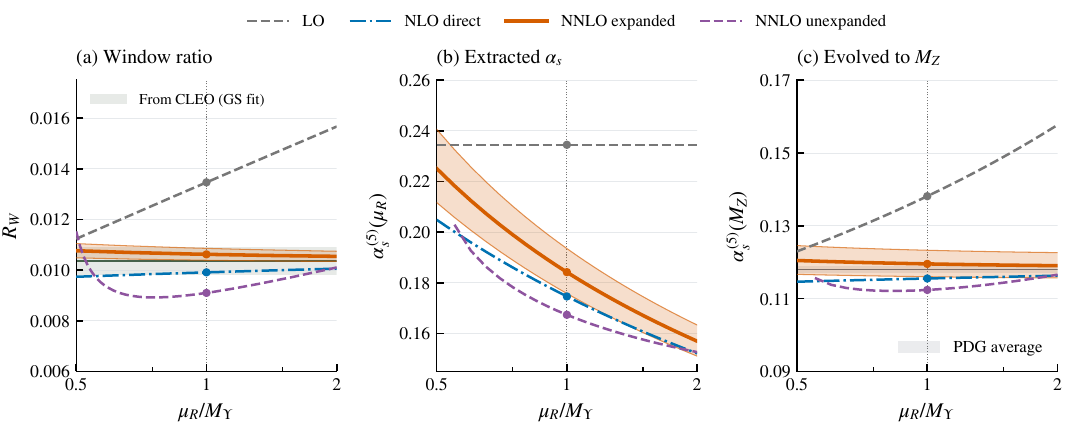}
\caption{Scale dependence of (a) the window ratio $R_W$ at
$\alpha_s(M_\Upsilon)=\ReferenceAlpha$ and (b,c) $\alpha_s$ extracted from
CLEO's GS window ratio.
Shaded bands on expanded NNLO show numerical uncertainties in (a) and
experimental plus numerical uncertainties in (b,c).
The unexpanded curves in (b,c) start at
$\mu_R/M_\Upsilon=\UnexpandedScaleMinimum$, below which no physical solution
exists.}
\label{fig:ratio-alpha}
\end{figure*}

\Needspace*{3\baselineskip}
\textit{Hadronic decay.---}
With the same inputs, the inclusive hadronic widths are
\begin{equation}
 \begin{aligned}
 \Gamma_h^{\rm LO}&=\HadronWidthLO\ {\rm keV},\\
 \Gamma_h^{\rm NLO}&=(\HadronWidthNLO\pm\HadronWidthNLOError_{\rm MC})\ {\rm keV},\\
 \Gamma_h^{\rm NNLO}&=(\HadronWidthNNLO\pm\HadronWidthMC_{\rm MC}
                    \pm\HadronWidthQuad_{\rm quad})\ {\rm keV},
 \end{aligned}
 \label{eq:hadronic-width}
\end{equation}
with central short-distance coefficients
\begin{equation}
 \begin{aligned}
 \bar c_h^{(1)}&=\HadronCOne\pm\HadronCOneError_{\rm MC},\\
 \bar c_h^{(2)}&=\HadronCTwo\pm\HadronCTwoMC_{\rm MC}\pm\HadronCTwoQuad_{\rm quad},
 \end{aligned}
 \label{eq:hadronic-coefficient}
\end{equation}
where $\mathrm{quad}$ denotes the two-loop quadrature error.
Converting to the four-flavor decoupling scheme \cite{DecouplingCKS}
reproduces the established NLO coefficient \cite{MackenzieLepage,SangFengJia}.

From the PDG $\Upsilon(1S)\to ggg$ branching fraction $(81.7\pm 0.7)\%$ and
total width $(54.02\pm 1.25)$~keV \cite{PDG}, we obtain
\begin{equation}
 \Gamma_h^{\rm PDG}=(\HadronPDGWidth\pm\HadronPDGWidthError)\ {\rm keV}.
 \label{eq:hadronic-reference}
\end{equation}
At $\mu_R=M_\Upsilon$ with our inputs, the positive first-order correction
raises $\Gamma_h^{\rm NLO}$ well above the measured total $\Upsilon(1S)$ width,
whereas the large negative second-order correction ($\bar c_h^{(2)}=\HadronCTwo$)
reduces $\Gamma_h^{\rm NLO}$ by $32.5\%$, bringing $\Gamma_h^{\rm NNLO}$ into
good agreement with $\Gamma_h^{\rm PDG}$ within uncertainties
[Fig.~\ref{fig:widths}(b)].
For $\mu_R\in\allowbreak[M_\Upsilon/2,2M_\Upsilon]$, $\Gamma_h^{\rm NNLO}$
varies by $-\HadronScaleDown\%/+\HadronScaleUp\%$ [Fig.~\ref{fig:widths}(b)],
as this negative second-order term grows rapidly at small $\mu_R$.

\textit{Window ratio and determination of $\alpha_s$.---}
Although the NNLO corrections bring both central widths close to experiment,
each width depends on $\langle O_1\rangle_\Upsilon$ and retains substantial
$\mu_R$ dependence, motivating the partial-width ratio
\begin{equation}
 R_W^{\rm th}(\alpha_s)=
 \frac{\Gamma_{\gamma,W}^{\rm th}(\alpha_s)}{\Gamma_h^{\rm th}(\alpha_s)},
 \label{eq:window-fit}
\end{equation}
in which the leading NRQCD matrix element $\langle O_1\rangle_\Upsilon$ and
kinematic mass factors cancel identically.
Because the Born widths scale as $\alpha\alpha_s^2$ and $\alpha_s^3$,
respectively, $R_W\propto\alpha/\alpha_s$ at leading order, allowing a direct
determination of $\alpha_s$.

Expanding $R_W^{\rm th}$ consistently through NNLO gives
\begin{equation}
 R_W^{\rm NNLO,exp}=R_W^{(0)}\biggl[1
 +c_{R,W}^{(1)}\frac{\alpha_s}{\pi}
 +c_{R,W}^{(2)}\left(\frac{\alpha_s}{\pi}\right)^2\biggr],
 \label{eq:ratio-expanded}
\end{equation}
where $R_W^{(0)}=\Gamma_{\gamma,W}^{(0)}/\Gamma_h^{(0)}$ and
\begin{align}
 c_{R,W}^{(1)}={}&c_{\gamma,W}^{(1)}-c_h^{(1)},\notag\\*
 c_{R,W}^{(2)}={}&c_{\gamma,W}^{(2)}-c_h^{(2)}
 +c_h^{(1)}\bigl(c_h^{(1)}-c_{\gamma,W}^{(1)}\bigr),
 \label{eq:ratio-coefficients}
\end{align}
with $c_{\gamma,W}^{(1)}$ including the $\mathcal O(\alpha\alpha_s^3)$
$C_g\otimes\allowbreak D_{g\to\gamma}$ term (replaced by
$c_{\gamma,W,\mathrm{dir}}^{(1)}$ at direct NLO).
At fixed $D_{i\to\gamma}$, substituting Eq.~\eqref{eq:scale-coefficients} into
Eq.~\eqref{eq:ratio-coefficients} cancels the quadratic logarithms $L_R^2$ and
the NRQCD matching logarithms $L_\Lambda$ identically, leaving
\begin{equation}
 \begin{aligned}
 c_{R,W}^{(1)}&=\bar c_{R,W}^{(1)}-\frac{\beta_0}{4}L_R,\\
 c_{R,W}^{(2)}&=\bar c_{R,W}^{(2)}-\frac{\beta_1}{16}L_R,
 \end{aligned}
 \label{eq:ratio-scale}
\end{equation}
whose linear logarithms compensate the running of the Born ratio through
relative $\mathcal O(\alpha_s^2)$.

\begin{table*}[t]
\caption{Strong coupling from CLEO's GS window ratio at $\mu_R=\mu_F=M_\Upsilon$,
with experimental (Data), numerical (Num.), and
$\mu_R\in[M_\Upsilon/2,2M_\Upsilon]$ (Scale) errors.
Dashes denote absent LO numerical error or no physical solution at
$\mu_R=M_\Upsilon/2$.}
\label{tab:alpha}
\centering
\renewcommand{\arraystretch}{1.18}
\setlength{\tabcolsep}{3pt}
\begin{tabular*}{\textwidth}{@{\extracolsep{\fill}}lcccccccc@{}}
\toprule
 & \multicolumn{4}{c}{$\alpha_s^{(5)}(M_\Upsilon)$}
 & \multicolumn{4}{c}{$\alpha_s^{(5)}(M_Z)$}\\
\cmidrule(lr){2-5}\cmidrule(lr){6-9}
Prediction & Central & Data & Num. & Scale & Central & Data & Num. & Scale\\
\midrule
\FitRows
\bottomrule
\end{tabular*}
\end{table*}

Numerically, the large second-order coefficients of
Eqs.~\eqref{eq:radiative-coefficient} and~\eqref{eq:hadronic-coefficient} also
nearly cancel in
$\bar c_{\gamma,W}^{(2)}-\allowbreak\bar c_h^{(2)}\simeq\allowbreak 1.7$,
yielding a moderate net coefficient $\bar c_{R,W}^{(2)}\simeq 15.7$.
At $\alpha_s(M_\Upsilon)=\allowbreak\ReferenceAlpha$, the perturbative series
reads
\begin{equation}
 \frac{R_W^{\rm NNLO,exp}}{R_W^{(0)}}
 =1-\RatioSeriesNLO+\RatioSeriesGluon+\RatioSeriesNNLO,
 \label{eq:ratio-series}
\end{equation}
where $-\RatioSeriesNLO$, $+\RatioSeriesGluon$, and $+\RatioSeriesNNLO$ are the
direct first-order correction, the $\mathcal O(\alpha\alpha_s^3)$
$C_g\otimes\allowbreak D_{g\to\gamma}$ contribution, and the second-order
correction, respectively.
Varying $\mu_R\in\allowbreak[M_\Upsilon/2, 2M_\Upsilon]$ changes
$R_W^{\rm NNLO,exp}$ by only $-\RatioScaleDown\%/+\RatioScaleUp\%$
[Fig.~\ref{fig:ratio-alpha}(a)], greatly reducing the scale dependence
relative to either width alone.

The unexpanded quotient
$R_W^{\rm NNLO,quot}\equiv\allowbreak\Gamma_{\gamma,W}^{\rm NNLO}/\allowbreak\Gamma_h^{\rm NNLO}$
agrees with Eq.~\eqref{eq:ratio-expanded} through NNLO, whereas evaluating the
numerator and denominator through NNLO without reexpanding the ratio enhances
relative $\mathcal O(\alpha_s^3)$ terms, especially near
$\mu_R=\allowbreak M_\Upsilon/2$ where $\Gamma_h^{\rm NNLO}$ becomes small
[Fig.~\ref{fig:ratio-alpha}(a)].
At $\mu_R=M_\Upsilon$ and the reference coupling, the two prescriptions give
\begin{equation}
 \begin{aligned}
 R_W^{\rm NNLO,exp}&=\WindowRatio\pm\WindowRatioError_{\rm num},\\
 R_W^{\rm NNLO,quot}&=\WindowQuotient\pm\WindowQuotientError_{\rm num}.
 \end{aligned}
 \label{eq:window-ratios}
\end{equation}

CLEO folded detector efficiency and resolution into the GS template and fitted
its normalization to the count spectrum in $W=[0.65,0.92]$, obtaining the
extrapolated total ratio
$R_{\gamma,\rm GS}^{\rm CLEO}=\allowbreak 0.0245\pm\allowbreak 0.0001_{\rm stat}\pm\allowbreak 0.0013_{\rm syst}$
\cite{CLEO2006,RadiativeAlpha2007}.
Multiplying by the reconstructed GS window fraction $K_{\rm GS}$
\cite{Garcia2005} gives the experimental window ratio:
\begin{equation}
 \begin{aligned}
 K_{\rm GS}&=\frac{\int_W S_{\rm GS}(z)\dd z}
 {\int_0^1 S_{\rm GS}(z)\dd z}=\GSFraction,\\
 \widehat R_W^{\rm GS}&=K_{\rm GS}R_{\gamma,\rm GS}^{\rm CLEO}
 =\GSWindowInput\pm\GSWindowDataError_{\rm data}.
 \end{aligned}
 \label{eq:window-input}
\end{equation}

Equating the expanded ratio $R_W^{\rm NNLO,exp}(\alpha_s)$ with
$\widehat R_W^{\rm GS}$ at the central scale $\mu_R=M_\Upsilon$ and evolving
the coupling to $M_Z$ at three loops yields (Table~\ref{tab:alpha})
\begin{equation}
 \begin{aligned}
 \alpha_s^{(5)}(M_\Upsilon)&=\FitExpandedLow
 {}^{+\FitExpandedLowDataPlus}_{-\FitExpandedLowDataMinus}
 {}^{+\FitExpandedLowNumericalPlus}_{-\FitExpandedLowNumericalMinus}
 {}^{+\FitExpandedLowScalePlus}_{-\FitExpandedLowScaleMinus},\\
 \alpha_s^{(5)}(M_Z)&=\FitExpandedHigh
 {}^{+\FitExpandedHighDataPlus}_{-\FitExpandedHighDataMinus}
 {}^{+\FitExpandedHighNumericalPlus}_{-\FitExpandedHighNumericalMinus}
 {}^{+\FitExpandedHighScalePlus}_{-\FitExpandedHighScaleMinus}.
 \end{aligned}
 \label{eq:alphas}
\end{equation}
The CLEO uncertainty dominates over both the numerical and scale uncertainties
in Eq.~\eqref{eq:alphas}.
Varying the GS reconstruction and endpoint extrapolation shifts
$\alpha_s(M_Z)$ by only $-\ReconstructionHighMinus$ to
$+\ReconstructionHighPlus$, while switching BFG sets or varying
$\mu_F\in\allowbreak[M_\Upsilon/2, 2M_\Upsilon]$ changes $\alpha_s(M_Z)$ by less
than $\FragmentationBoundUnits\times 10^{-5}$.

The expanded NNLO result agrees well with the world average $0.1180\pm 0.0009$
\cite{PDG}.
By contrast, the unexpanded quotient gives $\alpha_s(M_Z)=\FitQuotientHigh$
[Fig.~\ref{fig:ratio-alpha}(c)]; this $-\PrescriptionDifference$ shift far
exceeds the scale uncertainty of the expanded result in Table~\ref{tab:alpha}
and reflects sensitivity to uncalculated higher-order QCD terms, an issue also
seen in fixed-order $\alpha_s$ fits to low-energy $R(s)$ data
\cite{KataevTodyshev2026}.

\textit{Conclusions.---}
We have computed complete NNLO QCD corrections to radiative and inclusive
hadronic $\Upsilon(1S)$ decays in NRQCD at leading order in $v$.
Substantial negative NNLO corrections improve agreement with the CLEO photon
spectrum and partial width in $0.65\leq z\leq 0.92$, while reducing the
inclusive hadronic width to agree with the PDG value and resolving its NLO
excess at $\mu_R=M_\Upsilon$.

Taking the window ratio $R_W$ eliminates the leading NRQCD matrix element; in
its NNLO expansion, all quadratic scale logarithms $L_R^2$ cancel exactly while
the central second-order coefficients nearly cancel numerically, reducing the
$\mu_R$ dependence to the percent level.
Matching the expanded ratio to CLEO data gives
$\alpha_s^{(5)}(M_Z)=\FitExpandedHighWithError$, consistent with the world
average and limited within this fit by the CLEO uncertainty, whereas the
$-\PrescriptionDifference$ shift under the unexpanded quotient reflects
remaining sensitivity to higher perturbative orders.
Direct measurements of the photon spectrum and $R_W$ in a photon-energy window
at Belle~II without total-width extrapolation---combined with higher-order
perturbative, relativistic, and endpoint-resummed calculations---will improve
future $\alpha_s$ determinations and precision tests of heavy-quarkonium QCD.

\begin{acknowledgments}
Support for this work was provided by the National Natural Science
Foundation of China (Grant No.~12135013).
Generative artificial intelligence tools were used in developing code, checking
analytic and numerical results, and preparing the manuscript.
\end{acknowledgments}
\clubpenalty=150
\widowpenalty=150
\bibliography{references}
\end{document}

%% file: window_paper_data.tex
\newcommand{\InverseAlphaZero}{137.036}

\newcommand{\UpsilonMass}{9.4604}
\newcommand{\ReferenceAlpha}{0.18049}
\newcommand{\WindowWidthLO}{0.7831}
\newcommand{\WindowWidthNNLO}{0.4201}
\newcommand{\WindowWidthNLO}{0.7141}
\newcommand{\WindowWidthError}{0.0023}
\newcommand{\WindowWidthNLOError}{0.0016}

\newcommand{\WindowNNLOReduction}{41.2}

\newcommand{\WindowNLOScaleDown}{16.75}
\newcommand{\WindowNLOScaleUp}{0.14}

\newcommand{\WindowNNLOMinimum}{0.1295}
\newcommand{\WindowNNLOMaximum}{0.5483}
\newcommand{\WindowNNLOScaleDown}{69.18}
\newcommand{\WindowNNLOScaleUp}{30.52}
\newcommand{\ExtraSquarePercent}{0.31}
\newcommand{\ExtraInterferencePercent}{0.69}
\newcommand{\ExtraTotalPercent}{1.00}
\newcommand{\HadronCOne}{3.0576}
\newcommand{\HadronCTwo}{-115.72}
\newcommand{\HadronCTwoMC}{5.13}
\newcommand{\HadronCTwoQuad}{0.87}
\newcommand{\HadronWidthLO}{58.16}
\newcommand{\HadronWidthNLO}{68.381}
\newcommand{\HadronWidthNNLO}{46.17}

\newcommand{\HadronWidthNLOError}{0.015}
\newcommand{\HadronWidthMC}{0.99}
\newcommand{\HadronWidthQuad}{0.17}
\newcommand{\HadronCOneError}{0.0045}
\newcommand{\HadronPDGWidth}{44.13}
\newcommand{\HadronPDGWidthError}{1.09}

\newcommand{\HadronScaleDown}{75.67}
\newcommand{\HadronScaleUp}{17.53}

\newcommand{\WindowCOne}{-1.5341}
\newcommand{\WindowCOneError}{0.0365}
\newcommand{\WindowCTwo}{-113.99}
\newcommand{\WindowCTwoError}{1.03}
\newcommand{\RatioSeriesNLO}{0.2638}
\newcommand{\RatioSeriesGluon}{0.0009}
\newcommand{\RatioSeriesNNLO}{0.0519}
\newcommand{\RatioScaleDown}{0.75}
\newcommand{\RatioScaleUp}{1.34}
\newcommand{\GSFraction}{0.42306}
\newcommand{\GSWindowInput}{0.010365}
\newcommand{\GSWindowDataError}{0.000552}
\newcommand{\WindowRatio}{0.010622}
\newcommand{\WindowRatioError}{0.000235}
\newcommand{\WindowQuotient}{0.009100}
\newcommand{\WindowQuotientError}{0.000203}

\newcommand{\FitExpandedLow}{0.1842}
\newcommand{\FitExpandedLowDataMinus}{0.0078}
\newcommand{\FitExpandedLowDataPlus}{0.0085}
\newcommand{\FitExpandedLowNumericalMinus}{0.0035}
\newcommand{\FitExpandedLowNumericalPlus}{0.0036}
\newcommand{\FitExpandedLowScaleMinus}{0.0012}
\newcommand{\FitExpandedLowScalePlus}{0.0023}
\newcommand{\FitExpandedHigh}{0.1195}
\newcommand{\FitExpandedHighDataMinus}{0.0032}
\newcommand{\FitExpandedHighDataPlus}{0.0034}
\newcommand{\FitExpandedHighNumericalMinus}{0.0014}
\newcommand{\FitExpandedHighNumericalPlus}{0.0015}
\newcommand{\FitExpandedHighScaleMinus}{0.0005}
\newcommand{\FitExpandedHighScalePlus}{0.0009}

\newcommand{\FitQuotientHigh}{0.1124}

\newcommand{\FitRows}{LO & $0.2344$ & $\substack{+0.0132\\-0.0118}$ & \textemdash & $\substack{+0.0667\\-0.0417}$ & $0.1381$ & $\substack{+0.0043\\-0.0040}$ & \textemdash & $\substack{+0.0196\\-0.0152}$ \\
NLO direct & $0.1746$ & $\substack{+0.0072\\-0.0067}$ & $\substack{+0.0004\\-0.0004}$ & $\substack{+0.0018\\-0.0021}$ & $0.1155$ & $\substack{+0.0030\\-0.0029}$ & $\substack{+0.0002\\-0.0002}$ & $\substack{+0.0008\\-0.0009}$ \\
\textbf{NNLO expanded} & $\mathbf{0.1842}$ & $\substack{+0.0085\\-0.0078}$ & $\substack{+0.0036\\-0.0035}$ & $\substack{+0.0023\\-0.0012}$ & $\mathbf{0.1195}$ & $\substack{+0.0034\\-0.0032}$ & $\substack{+0.0015\\-0.0014}$ & $\substack{+0.0009\\-0.0005}$ \\
NNLO unexpanded & $0.1674$ & $\substack{+0.0056\\-0.0053}$ & $\substack{+0.0019\\-0.0019}$ & \textemdash & $0.1124$ & $\substack{+0.0024\\-0.0024}$ & $\substack{+0.0008\\-0.0008}$ & \textemdash \\}

\newcommand{\UnexpandedScaleMinimum}{0.55}

\newcommand{\ReconstructionHighMinus}{0.00034}
\newcommand{\ReconstructionHighPlus}{0.00045}

\newcommand{\FitExpandedHighTotalMinus}{0.0036}
\newcommand{\FitExpandedHighTotalPlus}{0.0038}
\newcommand{\FitExpandedHighWithError}{\FitExpandedHigh^{+\FitExpandedHighTotalPlus}_{-\FitExpandedHighTotalMinus}}

\newcommand{\FragmentationBoundUnits}{6}

\newcommand{\PrescriptionDifference}{0.0071}